\documentclass[a4paper,11pt]{article}
\usepackage{pos_adapted}

\usepackage{graphicx}
\usepackage{subcaption}

\newcommand{\mcut}{M_{\rm cut}}

\title{Spin Correlations in $t\bar{t}$ Production and Decay at the LHC
  in Perturbative QCD}

\ShortTitle{$t\bar{t}$ production and decay in perturbative QCD}

\author*[a,b,1]{Emanuele Re}

\affiliation[a]{Dipartimento di Fisica G. Occhialini, Universit\`a degli Studi di Milano-Bicocca}
\affiliation[b]{INFN, Sezione di Milano-Bicocca, Piazza della Scienza 3,20126 Milano, Italy}

\note{With minor differences, it matches the submitted proceedings for talks given at
  \begin{itemize}
  \item[-] \emph{The 33rd International Workshop on Deep Inelastic
  Scattering and Related Subjects (DIS2026)}, 4 - 8 May 2026, Bologna,
    Italy.
  \item[-] \emph{The XII edition of QCD@Work, International Workshop
  on QCD, theory and experiment (QCD@Work 2026)}, 16 - 19 June 2026,
    Trani, Italy.
  \end{itemize}
}

\emailAdd{emanuele.re@mib.infn.it}

\abstract{

  Among recent developments in top-quark physics at the LHC, the
  production of top-quark pairs near threshold has attracted
  considerable attention, partly due to tensions between data and
  theoretical predictions.
  We show that reliable predictions for threshold-sensitive
  observables, such as spin correlations and the differential
  distribution in the \(t{\bar t}\) invariant mass, can be obtained in
  perturbation theory by retaining the first few powers of
  \(\alpha_s/v\), where \(v\) is the relative velocity of the top
  quarks. Applying this approach to reweight nominal Monte Carlo
  predictions for spin-sensitive observables, we find good agreement
  with data. We also discuss how threshold effects, including
  bound-state contributions, can be incorporated on an event-by-event
  basis into fully differential Monte Carlo event generators.
}

\begin{document}
\maketitle

\section{Introduction}
\label{sec:intro}
Top-pair production at the LHC is generally described well by
fixed-order QCD calculations, possibly augmented with different types
of resummation, depending on the observable at hand. Fully exclusive
predictions at (N)NLO matched to parton showers also exist and are, in
general, successful in capturing typical kinematical features. A
special situation arises, however, when the invariant mass of the pair
approaches the production threshold, $M_{t\bar t}\simeq 2 m$, $m$
being the top-quark mass. The top and antitop are then produced with a
small relative velocity \(v\), and gluon exchange between them generates
Coulomb corrections enhanced by inverse powers of \(v\). The
perturbative expansion consequently contains terms of the form
$({\alpha_s}/{v})^n$. For \(v\sim\alpha_s\), all orders become
parametrically important, and a non-relativistic resummation is
formally required.

This issue has become particularly relevant in recent years, in
connection with measurements of \(t\bar t\) production at the LHC.
Indeed, the tensions observed by ATLAS and CMS between data and theory
for spin correlations and cross sections close to threshold have
naturally prompted discussions of the possible production of the
pseudoscalar \(t\bar t\) bound state, usually denoted \(\eta_t\) and
referred to as ``toponium''.

In refs.~\cite{Nason:2025hix,Nason:2026oka} we have studied these
aspects in detail. In ref.~\cite{Nason:2025hix} we have shown that the
observables relevant for existing spin-correlation measurements are
sufficiently inclusive that the dominant threshold effects can be
treated perturbatively, retaining only the first few enhanced
terms. In ref.~\cite{Nason:2026oka}, which appeared after this talk was
given, we instead focus on differential cross sections rather than
spin correlations, implementing threshold-enhanced corrections in
NLO+PS generators, including finite-width effects, with some
approximations.

Our main finding is that threshold enhancement, notably that due to
the formation of a bound state, can be computed using genuine
perturbative QCD, as long as measurements are sufficiently inclusive,
i.e.~if the experimental resolution on $M_{t\bar t}$ is not good
enough to resolve the peak of toponium production. We have also found
that care must be taken when adding the bound-state contribution on
top of the so-called ``continuum'' production, as the two are
intrinsically intertwined.

\section{Threshold dynamics and spin correlations}
\label{sec:spin}
Near threshold, the \(t\bar t\) system is non-relativistic. At leading
order, the phase-space factor is proportional to \(v\), while Coulomb
gluon exchange produces, above threshold, the Sommerfeld enhancement,

$$
F(z)=\frac{z}{1-e^{-z}},
\qquad
z=\frac{\pi a_l}{v},
$$

where \(l\) denotes the colour representation of the final state. For
the singlet and octet channels, one has $a_1= C_F\alpha_s,
a_8=-\frac{1}{2N_c}\alpha_s$.
Expanding the Sommerfeld factor gives
$F(z)=1+\frac{z}{2}+\frac{z^2}{12} -\frac{z^4}{720}+\cdots$, thus the
first two threshold-enhanced corrections scale as $\frac{\alpha_s}{v}$
and $\frac{\alpha_s^2}{v^2}$.

The colour-singlet channel ($l=1$) is particularly interesting because
the Coulomb potential $V(r)=-a_l/r$ is attractive: for this reason,
bound states can in principle form. Indeed, as discussed later, the
all-order spectral density for colour-singlet production contains
bound-state poles. The octet potential, instead, is repulsive, and no
bound states are present.

The production dynamics at threshold also has direct implications for
spin correlations: since threshold production is dominated by
\(S\)-wave configurations,\footnote{For bound states, this can be
understood by recalling that, in ordinary quantum mechanics, the
spatial wave function for a hydrogen-like atom scales as $p^\ell$,
where $p$ is the relative momentum of the two particles and $\ell$ is
the orbital quantum number.} the total angular momentum of the final
state is essentially determined by the \(t\bar{t}\) spin state, which,
in turn, is related to the production channel. The \(gg\) channel, which
has both colour-singlet and colour-octet components, is dominated by
spin-singlet production, corresponding to pseudoscalar quantum numbers
for the $t\bar{t}$ final state. The \(q\bar{q}\) channel, instead, has
only a colour-octet component and is dominated by spin-triplet
production. Since \(gg\) production is the dominant production channel
at the LHC, it is easy to see, by estimating the NLO threshold
corrections to its two colour components (section 2 of
ref.~\cite{Nason:2025hix}), that the leading threshold correction to
$t\bar{t}$ production is positive.

The charged leptons from top decay provide efficient spin analyzers.
The typical observable used to study spin correlations is the
dilepton observable
$
C_{\rm hel}=\hat l_1\cdot\hat l_2,
$
where the lepton directions are defined in the respective top and
antitop rest frames, after first boosting the lepton momenta to the
$t\bar{t}$ rest frame. In the ideal threshold limit, the spin-singlet
configuration dominates, and one obtains
$
\frac{1}{\sigma}\frac{d\sigma}{dC_{\rm hel}}
=\frac{1+C_{\rm hel}}{2},
$
and therefore
$
\langle C_{\rm hel}\rangle=\frac13.
$
The observable used in the experimental analyses can equivalently be
expressed in terms of $D=-3\langle C_{\rm hel}\rangle$, which is one of
the markers of entanglement~\cite{Afik:2022kwm}.

\section{Perturbative treatment of the threshold region}
\label{sec:perturbative}
In ref.~\cite{Nason:2025hix} we focused on the binned differential
cross section $d\sigma/dM_{t\bar t}$ close to threshold and on the
observable $D$. Such quantities are effectively obtained by
integrating the threshold-enhanced spectrum up to an upper
invariant-mass cut $M_{cut}$. For the phenomenological reasons
mentioned above (and because of the obvious limitations in the
experimental resolution on $M$), this cut is not much larger than
$2 m$, although it is at least 5--10 GeV higher. In this scenario,
not only can Coulomb effects be expanded perturbatively, but the
contribution from the formation of quasi-bound states can also be
estimated purely using perturbation theory. In the rest of this
section, I will explain the main ideas.

First of all, the characteristic relative velocity $v_{cut}$ of the
top quark in the $t\bar{t}$ rest frame is set by the kinematic range
of the observable: if $M_{t\bar{t}}<M_{cut}$, then
$v_{cut}=\sqrt{(M_{cut}-2 m)/m}$, since in the $t\bar{t}$ rest frame,
$M_{t\bar{t}} = E_{t\bar{t}}\simeq 2 m + m v^2$. Schematically,
considering only threshold-enhanced terms and defining ${\cal
  E}=M_{t\bar{t}}-2 m$, one has
$$
\sigma(M_{t\bar{t}}<M_{cut}) \sim
\int^{{\cal E}_{cut}}\!\!\!\!\!\!\!\! d{\cal E}\ \sigma(E_{t\bar{t}})
\sim \frac{\alpha_s^2}{m}\ v_{cut}^3\
\left(1 + c_{1,l} (a_l/v_{cut}) + c_{2,l} (a_l/v_{cut})^2 + ...\right)
$$
where, with a slight abuse of notation, we use the same equation to
represent the expansion parameters appropriate to the colour-singlet
and colour-octet channels. We have extracted an overall factor
$\alpha_s^2$, since we are computing corrections relative to the LO
hard $t\bar{t}$ partonic cross section. The cubic $v^3_{cut}$
dependence arises because $\sigma(E)\sim v$ (due to the $t\bar{t}$
phase space), while ${\cal E}\sim m v^2$.

It is not difficult to see that bound-state contributions are of
relative order $a_l^3$ and, obviously, contribute only if $l=1$. In
fact, the description of the lowest-energy $t\bar{t}$ quasi-bound
state parallels that of a positronium system in ordinary quantum
mechanics. The size of the $t\bar{t}$ system is characterized by the
Bohr radius $r_b=2/(m a_1)$, hence the wave function of the system at
the origin must scale as $|\psi(r=0)|^2\sim (1/r_b)^3$. Therefore,
bound states contribute to the cross section $\sigma$ as
\begin{equation}
\label{eq:psi0squared}
\sigma(E_{t\bar{t}}) \sim \frac{\alpha_s^2}{m^4}\
\delta(E_{t\bar{t}}-2 m)|\psi(r=0)|^2
\sim\frac{\alpha_s^2}{m}\ \delta(E_{t\bar{t}}-2 m)\ a^3_1\,,
\end{equation}
hence, from the point of view of standard perturbation theory,
bound-state effects in $\sigma$ arise at order N$^3$LO, even though
such contributions cannot appear in a purely perturbative
computation, since a perturbative series is analytic in $\alpha_s$,
whereas the bound-state contribution is absent for the colour-octet
configuration, i.e.~when $a_l$ is negative ($l=8$).

On the other hand, in ref.~\cite{Nason:2025hix} we have shown, through
a simplified argument followed by a formal proof, that the cumulative
distribution $\int^{{\cal E}_{cut}}\! d{\cal E}\ \sigma(E)$ can be
computed in perturbation theory, properly taking into account the
bound-state contributions. In summary, recalling that the spectral
density $\rho(\cal{E})$ contains both continuum and bound-state
contributions, and that its integral is proportional to the cross
section, it can be shown that
\begin{itemize}
\item upon integrating the spectral density over a sufficiently wide
  interval (i.e.~with ${\cal E}_{cut}$ sufficiently above $0$), the
  non-analytic term arising from bound states combines with a similar
  term from the continuum, yielding an analytic result for the cross
  section;
\item as long as the coefficients of the perturbative expansion of
  $\rho(\cal{E})$ are interpreted distributionally, integrating the
  perturbatively expanded spectral density over energy reproduces the
  result obtained by expanding the integrated expression.
\end{itemize}
In equations:
\begin{eqnarray}
  \rho_l(\cal{E}) & = &
  \underbrace{\theta(a_l) \frac{1}{\pi r_b^3}
  \sum_{n=1}^\infty \frac{1}{n^3} \delta({\cal E}-E_{l,n})}_{bound}
  +
  \underbrace{\frac{m^{3/2}}{4\pi^2} \sqrt{{\cal E}}
  F(\pi a_lv^{-1})}_{cont.}\,, \label{eq:rho1} \\
  \rho_l(\cal{E}) & \to &
  \frac{1}{2} \frac{1}{\pi r_b^3}
  \sum_{n=1}^\infty \frac{1}{n^3} \delta({\cal E}-E_{l,n})
  +\frac{m^{3/2}}{4\pi^2} \sqrt{{\cal E}}
  F^+(\pi a_lv^{-1})\,, \label{eq:rho2}
\end{eqnarray}
where integrating the perturbative expansion of eq.~\ref{eq:rho2}
yields, order by order, the same result as integrating
eq.~\ref{eq:rho1}. Notice that, as expected, the leading
contribution from bound states scales as $\alpha_s^3$ relative to
the leading-order production cross section, as in
eq.~\ref{eq:psi0squared}. Moreover, besides the fact that
eq.~\ref{eq:rho2} is now purely analytic in $\alpha_s$, it is
crucial to notice the factor $1/2$: the expansion of the integral of
the continuum contains a term of order $\alpha_s^3$ that corrects
the contribution from the purely bound-state part. This is the
interplay mentioned in Sec.~\ref{sec:intro}: the contribution that
one might naively identify with the bound state is reduced by a
factor of two once the complete perturbative expansion is considered.
Among other things, this fact needs to be taken into account if one
attempts to model the ``toponium'' contribution by simply adding it
to the continuum result obtained using perturbation theory.
\footnote{The observation that bound-state contributions should not
be added to cross sections involving integrated spectral densities is
not new: see refs.~\cite{Beneke:2016jpx, Melnikov:2014lwa,
Eides:2014swa, Braun:1968njz,Novikov:1977dq, Voloshin:1979uv,
Voloshin:1984zzn, Smith:1994ev}.}


\section{Predictions at Threshold at the LHC}
\label{sec:paper1}
For the observables studied in ref.~\cite{Nason:2025hix}, we have
included the first few threshold-enhanced terms of eq.~\ref{eq:rho2},
$$
\rho_l({\cal E}) \simeq  \frac{m^2}{4\pi^2}\left(v+\frac{b_l}{2} +\frac{b_l^2}{12v}
  +\frac{\zeta(3)}{4\pi^2} b_l^3 m \delta({\cal E})
  + {\cal O}(\alpha_s^4) \right)\,, \ \ \ b_l=\pi a_l\,, \label{eq:rhottexpanded}
$$
by correcting standard Monte Carlo predictions through the subtraction
of terms already present at the corresponding fixed order. This
provides a consistent matching between ordinary perturbative QCD and
the threshold expansion.\footnote{The scale entering the Coulomb
coupling is another important issue. It should not be chosen simply as
the arbitrarily small top-quark momentum of an event infinitesimally
above threshold. For an integrated observable, the relevant scale is
instead related to the characteristic momentum associated with the
upper boundary of the integration region. This leads to a perturbative
scale of order several tens of GeV for the cuts used in the LHC
analyses.}

\begin{figure}[htb]
  \begin{center}
    \includegraphics[width=0.45\textwidth]{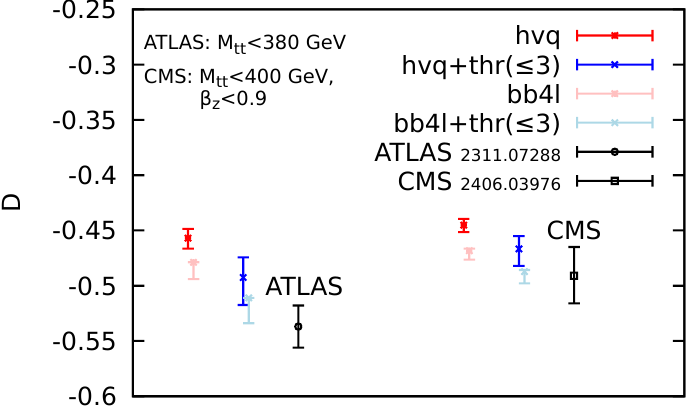}\hfill
    \includegraphics[width=0.5\textwidth]{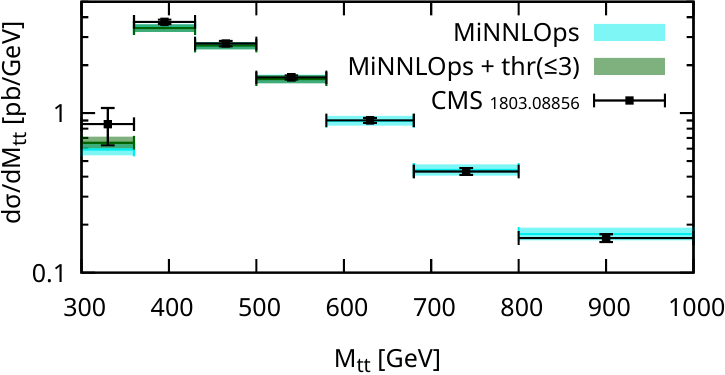}
  \end{center}
  \caption{\label{fig:paper1} Results from
    ref.~\cite{Nason:2025hix}. Left: Prediction for the $D$ observable
    as measured by ATLAS and CMS, obtained with the nominal POWHEG
    {\tt hvq} and {\tt bb4l} generators, and with the inclusion of
    enhanced threshold corrections up to N$^3$LO. Right:
    Invariant-mass distribution of the $t\bar{t}$ pair compared with
    CMS data. The MiNNLO predictions are shown with and without the
    inclusion of threshold corrections up to the cubic term.}
\end{figure}

The threshold corrections were included by suitably modifying
distributions obtained using NLO+PS (and NNLO+PS) generators. The
results show that the enhanced corrections have a visible impact on
the predicted spin correlations when a low \(M_{t\bar t}\) cut is
imposed. As shown in the left panel of fig.~\ref{fig:paper1}, adding
the leading threshold-enhanced terms shifts the NLO(+PS) prediction
for the spin-correlation observable \(D\) towards the experimental
data. As expected, the effect is reduced when correcting NNLO(+PS)
results (not shown in the figure), because the leading
threshold-enhanced terms are already included in the higher-order
fixed-order calculation.

In ref.~\cite{Nason:2025hix} we have also computed predictions for the
binned differential cross section $d\sigma/dM_{t\bar t}$ close to
threshold. We found that including threshold-enhanced terms at
${\cal O}(\alpha_s^3)$ (where bound-state effects enter, as
explained previously) is useful, but not crucial, for describing the
CMS data of ref.~\cite{CMS:2018htd}, once NNLO results are considered
(right panel of fig.~\ref{fig:paper1}). It is important to note that
this conclusion might change to some extent for more exclusive
$M_{t\bar t}$ measurements, such as those of
refs.~\cite{CMS:2025kzt,ATLAS:2026dbe}.

The study of ref.~\cite{Nason:2025hix} also demonstrates that the
treatment of top-quark decays matters. Since the charged leptons
analyse the top spins, approximations in the treatment of the decays
can affect the predicted correlations.

\section{Off-shell effects in NLO+PS description of threshold production}
\label{sec:paper2}

As summarized in the previous section, in ref.~\cite{Nason:2025hix}
threshold corrections were included only at the histogram level and
expanded only up to N$^3$LO.
In ref.~\cite{Nason:2026oka}, we extended these results by
incorporating all these subleading effects into event generators. This
is particularly important because event generators are essential at
every stage of experimental analyses and play a central role in the
ATLAS and CMS studies, such as those discussed in
Sec.~\ref{sec:intro}, where they are used to establish whether an
excess over the SM prediction is present. In such cases, incorporating
threshold effects, including bound-state contributions, into event
generators is essential for a meaningful comparison between data and
theoretical predictions. The inclusion of off-shell effects is also
needed for a realistic simulation; ref.~\cite{Nason:2026oka} addresses
this issue as well.

A first point we discussed in ref.~\cite{Nason:2026oka} is that the
formal need for resummation close to threshold, i.e.~when
\(v\sim\alpha_s\), does not imply that every experimentally relevant
observable requires a fully resummed calculation. This is particularly
true for observables integrated over a sizable interval in \(M_{t\bar
  t}\). We have explicitly shown that, for the observables of interest
at the LHC, where ${\cal E}_{cut}>5$ GeV, including a few terms in the
expansion of the Sommerfeld factor in eq.~\ref{eq:rho2} gives a
reliable result. In other words, the all-order series converges quite
rapidly: the first Coulomb correction is large, and the second is
still sizable, but subsequent terms become progressively smaller. The
all-order result is consequently approached after including only the
first few enhanced contributions.

The main result of ref.~\cite{Nason:2026oka} is the construction of
three NLO+PS event generators that include threshold-enhanced
effects. These implementations allow a direct comparison with data at
the event-generation level, rather than only at the level of an
inclusive fixed-order cross section.

On the formal side, we have discussed how the spectral density is
modified when off-shell effects for the two top quarks are included.
Besides recovering the known results of Fadin and
Khoze~\cite{Fadin:1987wz,Fadin:1990wx}, we have also obtained an
equation that, in principle, allows for an exact implementation of
off-shell effects in event generators. Equivalent results were
already obtained in ref.~\cite{Bach:2017ggt} in the context of
$e^+e^-$ collisions.

By making approximations to the aforementioned equations, we have
included threshold effects in POWHEG NLO+PS event generators,
retaining finite-top-width effects and including bound-state
contributions as well.

We have upgraded the POWHEG {\tt hvq} code in two ways:
\begin{itemize}
  \item The implementation denoted {\tt thr1} is based on the standard
    POWHEG description of \(t\bar t\) production and incorporates
    threshold corrections in the narrow-width approximation. Off-shell
    effects are included only through the customary POWHEG
    reshuffling of final-state momenta, performed after the events
    have been generated.
  \item The second implementation, denoted {\tt thr2}, allows the top
    and antitop to be off shell. The fixed-order matrix elements are
    evaluated through a projection onto the corresponding on-shell
    configuration, while the full phase space allows the inclusion
    of the effects encoded in the off-shell spectral density. This
    provides a practical way of combining the enhanced production
    dynamics with finite-width effects.
\end{itemize}

The original {\tt bb4l} code provides a full off-shell description,
including the relevant resonant and non-resonant contributions and
spin correlations in the decays. It was improved by including
threshold corrections using a strategy similar to that of {\tt thr2},
with the top momenta defined such that the radiated gluon at NLO
contributes to the top momentum only when it is sufficiently close to
an outgoing $b$-quark.

The comparison between {\tt hvq-thr1}, {\tt hvq-thr2} and the improved
{\tt bb4l} calculations also provides a useful estimate of the
theoretical uncertainty associated with the narrow-width
approximation.

In the left panel of fig.~\ref{fig:paper2}, we show an example of the
differences obtained when binning the \(M_{t\bar t}\) distribution.
The right panel of fig.~\ref{fig:paper2}, instead, shows a comparison
between the {\tt thr2} results and those obtained in
ref.~\cite{Garzelli:2026ctb}.

\begin{figure}[htb]
  \begin{center}
    \begin{subfigure}[t]{0.50\textwidth}
      \vspace{0pt}
      \includegraphics[width=\textwidth,page=8]{paperPlots}
    \end{subfigure}
    \hfill
    \begin{subfigure}[t]{0.45\textwidth}
      \vspace{-12pt}
    {\scriptsize
      \begin{tabular}{|l||c||c||c|c|}
        \hline
        \multicolumn{5}{|c|}{{\tt thr2} NLO results, widths corr., 340~GeV$<m_{t{\bar t}}<\mcut$ } \\
        \hline
        $\mcut$ & $\Delta$ from \cite{Garzelli:2026ctb} & Baseline & central &  $\Delta$  \\
        \hline 
        345 & $ 2.73\pm 0.61 $ & 0.925(6) & 2.540(9) & 1.62(1)   \\
        \hline 
        346 & $ 3.16\pm 0.75 $ & 1.366(7) & 3.58(1) & 2.22(1)    \\
        \hline 
        347 & $ 3.48\pm 0.90 $ & 2.056(7) & 4.80(1) & 2.75(1)    \\
        \hline 
        348 & $ 3.74\pm 1.08 $ & 3.005(8) & 6.19(1) & 3.18(1)    \\
        \hline 
        349 & $ 3.97\pm 1.27 $ & 4.175(9) & 7.73(1) & 3.55(2)    \\
        \hline 
        350 & $ 4.15\pm 1.47 $ & 5.53(1) & 9.39(1) & 3.87(2)     \\
        \hline 
        351 & $ 4.31\pm 1.69 $ & 7.04(1) & 11.19(1) & 4.15(2)    \\
        \hline 
        352 & $ 4.44\pm 1.92 $ & 8.69(1) & 13.10(2) & 4.41(2)    \\
        \hline 
        353 & $ 4.56\pm 2.15 $ & 10.47(1) & 15.11(2) & 4.64(2)   \\
        \hline 
        354 & $ 4.65\pm 2.4 $ & 12.36(1) & 17.22(2) & 4.86(2)    \\
        \hline 
        355 & $ 4.72\pm 2.66 $ & 14.35(1) & 19.41(2) & 5.06(2)   \\
        \hline 
      \end{tabular}
    }
    \end{subfigure}
  \end{center}
  \caption{\label{fig:paper2} Results from
    ref.~\cite{Nason:2026oka}. Left: Invariant mass distribution of
    the $t\bar{t}$ pair (left).  Right: Comparison of the {\tt thr2}
    generator ({\tt dt+st+nt}, running width effects included) against
    ref.~\cite{Garzelli:2026ctb}. The $\Delta$ represent the increase
    in cross sections when threshold corrections are added. Cross
    sections are expressed in pb.}
\end{figure}

Work is currently being done to formally improve on the approximations
used to implement threshold-corrections into POWHEG NLO+PS event generators.

A further direction for future work will be a more systematic
comparison with other recent results, such as GFRW
(refs.~\cite{Sumino:2010bv,Fuks:2021xje,Fuks:2024yjj}, NRQCD
(refs.~\cite{Garzelli:2024uhe,Garzelli:2026ctb}), and the recent\ {\tt
  Pythia8} implementation (ref.~\cite{Sjostrand:2026rcz}).

\acknowledgments Results presented here have been obtained in
collaboration with P.~Nason, L.~Rottoli and G.~Pelliccioli. The author
thanks the organizers of DIS 2026 and QCD@Work 2026 for facilitating his participation in the
two conferences.

\end{document}